\documentclass[12pt]{article}
\usepackage{amssymb}
\usepackage{amsmath}
\def\ee{{\rm e}}\def\ii{{\rm i}}
\def\beq{\begin{equation}}\def\eeq{\end{equation}}
\def\bea{\begin{eqnarray}}\def\eea{\end{eqnarray}}

\begin{document}

\title{A Geometrical Description of Spinor Fields: non-Grassmannian toy model}

\author{Roman Sverdlov\\
\\Physics Department, University of Michigan,
\\450 Church Street, Ann Arbor, MI 48109-1040, USA}
\date{August 04, 2008}
\maketitle

\begin{abstract}
\noindent In this paper I am going to present the way to define fermionic field in terms of three orthogonal vector fields of norm 1 together with two real valued scalar fields. This paper is based on a toy model where there are no Grassmann variables. 
\end{abstract}

\noindent{\bf 1. Introduction}

\noindent The goal of this paper is to come up with a geometric model to describe fermions, or more specifically spin-half Dirac spinors. The possibility of reformulating spinors geometrically is in general an interesting issue for quantum field theory, and it is certainly not a new idea (see, for example, Penrose's flagpole-plus-flag formulation \cite{Penrose}), but in the present case part of the motivational context is provided by my work on the dynamics of various types of fields coupled to gravity in the causal set approach \cite{causets}, and I seek a formulation in which the geometrical structure is suitable for translation into causal sets terms. For our description of the dynamics of gravity and coupled scalar fields in the causal set approach see Ref \cite{paperI}; For the use of results in this paper in the dynamics of spinors coupled to gravity in the causal set approach see Ref \cite{spinors}, and for a corresponding description of gauge theories see Ref \cite{gauge} and Ref \cite{bosonic}. 

This paper considers a toy model where there are no Grassmann variables. A complete definition of fermionic field that takes into account Grassmann variables is treated in Ref \cite{Grassmann}

\noindent{\bf 2. Setup}

\noindent If we have an arbitrary spinor at a point, we can always rotate it into a state of the form $\chi_p u_1 + \chi_a v_1$ (here "p" stands for particle, and "a" stands for antiparticle). This can be seen by counting degrees of freedom. The rotation group in 4 dimensions has 6 degrees of freedom, while multiplication by an arbitrary complex scalar adds 2 degrees of freedom. This means that if the actions of these two groups were independent, we would obtain a total of 8 real degrees of freedom, which matches the number of real degrees of freedom in a 4-spinor. 

In light of the above, I would like to get rid of the notion of fermion in favor of more ``geometrical" quantities, which are:

\noindent(1) 6 orthonormal vierbeins which define local frame in which spinor has a form $\chi_p u_1 + \chi_a v_1 $, where both $\chi_p$ and $\chi_a$ are real. 

\noindent(2) $\chi_p$ and $\chi_a$ (see above).

\noindent Thus, I will rotate the reference frame from point to point in such a way that the fermionic field is always in the desired form.

Now, on the event that we have more than one fermionic particle, we will have many different sets of vierbeins. For instance, if we have electron and neutrino, then we have one set of veirbeins that would put electron into a form $\chi_{ep} u_1 + \chi_{ea} v_1$ and the other set of veirbeins that would put neutrino into a form $\chi_{\nu p} u_1 + \chi_{\nu a} v_1$ . This should be contrasted with the way vierbeins are normally treated where there is only one veirbein that is intrinsic part of geometry. The reason for this is that right now vierbeins are no longer viewed as definition of geometric "stage" but rather as attributes of fermionic "actors", which means it would make sense that the more actors we have the more vierbeins we have as well. 

In order to stress the fact that vierbines are now viewed as fields, I will replace $e_0^{\mu}$, $e_1^{\mu}$, $e_2^{\mu}$ and $e_3^{\mu}$ by $A^{\mu}$, $B^{\mu}$, $C^{\mu}$ and $D^{\mu}$ respecively, and introduce Lagrange multipliers to assure that 

\bea
& &A^{\mu}A_{\mu}=1\;,\quad B^{\mu}B_{\mu} = C^{\mu}C_{\mu}
= D^{\mu}D_{\mu} = -1\;, \\
& &A^{\mu}B_{\mu} = A^{\mu}C_{\mu} = A^{\mu}D_{\mu}
= B^{\mu}C_{\mu} = B^{\mu}D_{\mu} = C^{\mu}D_{\mu} = 0
\eea
(I am using a metric of signature $(+,-,-,-)$). We will then relax the assumption about rotation of reference frames and go back to the flat Minkowski case. Thus, the final form for my notation for a spinor will be $(A^\mu, B^\mu, C^\mu, D^\mu, \phi, \chi)$.

The geometrical model I propose has also an intuitive appeal: if we take the word ÒspinÓ literally and imagine a particle spinning,  we would need to know the plane in which the particle spins. This gives us two axes, which are described by two vectors, $B^{\mu}$ and $C^{\mu}$. Now, since spin is subject to Lorentz transformations, we also need to know the rest frame of the particle, and this is determined by a timelike vector $A^{\mu}$. As far as $D^{\mu}$ is concerned, due to the orthogonality and unit-norm conditions, it is completely determined by the above 3 vectors. As you will see from the results of the ``Lagrangian" section, things can indeed be visualized in terms of spinning.

The biggest objection one can have is that vector fields and fermions have different transformation properties. However, if one realizes that the transformation properties are completely determined by the Lagrangians and inner products, we can cure the problem by drawing attention towards the latter two, and away from the transformation properties. For example, the implication of spin-$\frac12$ is that a $360^\circ$ rotation in vector space is the same as a $180^\circ$ rotation in spinor space. This problem can be cured by redefining what we mean by a rotation: Instead of simply using $U \mapsto M(\theta)\, U$, where $M(\theta)$ is the usual rotation matrix, we use $U \mapsto \ee^{\ii\theta/2}\, M(\theta)\, U$; by adding a phase, the complex amplitude switches sign upon a rotation by $\theta = 360^\circ$, despite the fact that the vectors are rotated back to their original positions. The reason for this feature is that SU(2) is not the full symmetry group; rather, the full symmetry group is ${\rm SU}(2) \times {\rm U}(1)$. This gives us the freedom of selecting a subgroup $R$ of ${\rm SU}(2) \times {\rm U}(1)$ such that $R \times {\rm U}(1) = {\rm SU}(2) \times {\rm U}(1)$. Any such $R$ can be used as a definition of rotation group, and the freedom of choosing this $R$ corresponds to a freedom in defining the value of the spin: spin-$\frac12\ \times$ spin-0 = spin-1 $\times$ spin-0. Another example: suppose we perform a $180^\circ$ rotation in the space of vectors. In this case, the fact that vectors determine a coordinate system doesn't stop us from {\em defining\/} the inner product between two flipped coordinate systems to be 0 instead of $-1$. After all, we can define the inner product any way we like, so we chose to do it this way. These two features will be implemented in the remainder of the paper.

Throughout this paper I will use the following representation:
$$\gamma^0 = \left( \begin{matrix}
{\bf1} & \hfill0\, \\ 0 & -{\bf1}\\ \end{matrix} \right),
\qquad
\gamma^k = \left( \begin{matrix}
0 & \sigma^k \\ -\sigma^k & 0 \end{matrix} \right).
$$
where $\bf 1$ is the $2\times2$ unit matrix, $\sigma^k$ for $k=1,2,3$ are the Pauli matrices, and the basis column state vectors will be defined as follows:
$$u_1 = \left( \begin{matrix}
1 \\ 
0 \\ 
0 \\ 
0 \end{matrix} \right)\;,\qquad
u_2 = \left( \begin{matrix}
0 \\ 
1 \\ 
0 \\ 
0 \end{matrix} \right)\;,\qquad
v_1 = \left( \begin{matrix} 
0 \\ 
0 \\ 
1 \\ 
0 \end{matrix} \right)\;,\qquad
v_2 = \left( \begin{matrix}
0 \\ 
0 \\ 
0 \\ 
1 \end{matrix} \right)\;.$$

\noindent{\bf 4. Free Lagrangian}

\noindent 

Even though in this paper we are only dealing with a toy model in which there are no Grassmann numbers, we are still free to get rid of $\chi_p^2$ and $\chi_a^2$ terms of the Lagrangian. This means that as far as spin connection terms are concerned, we are looking only at $\chi_p \chi_a$ terms. Based on the fact that spinors take the above form, it is apparent that the only term of $\omega_{mab} \overline{\psi} \gamma^m \sigma^{ab} \psi = \omega_{mab} \psi^{\dagger} \gamma^0 \gamma^m \sigma^{ab} \psi$ that survives is the one where $\gamma^0 \gamma^m \sigma^{ab}$ is off-diagonal matrix in the $2 \times 2$ block representation. This will happen only if m, a and b are all non-zero, which identifies them as $1$, $2$ and $3$ up to permutations, which means they are all proportioanl to $\psi^{\dagger} \gamma^5 \psi =2 \chi_p \chi_a$. 

As far as derivative terms, we do have to keep $\psi_p \partial \psi_p$ and $\psi_a \partial \psi_a$ terms as well as we still have to keep the "mixed" ones. This means that we are looking both at the diagonal and off diagonal matrices in block diagram. However, since there are no spin down components of either particle or antiparticle, each block needs to be diagonal. The matrices that satisfy these constraints are $\gamma^0$ and $\gamma^3$ . $\gamma^0$ will give us $\overline{\psi} \gamma^0 e^{0 \mu} \partial_{\mu} \psi = e^{0 \mu} (\chi_p \partial_{\mu} \chi_p + \chi_a \partial_{mu} \chi_a)$ and $\gamma^3$ will give us $\overline{\psi} \gamma^3 e^{3 \mu} \partial_{\mu} \psi = e^{3 \mu} (\chi_p \partial_{\mu} \chi_p - \chi_a \partial_{mu} \chi_a)$ Thus, the Lagrangian becomes 

\bea & & {\cal L}_{free}= k \psi^{\dagger} \gamma^5 \psi (\omega^1_{23}-\omega^2_{13}+\omega^3_{12}) + \overline{\psi} e^{0 \mu} \gamma_0 \partial_{\mu}  \psi +  \overline{\psi} e^{3 \mu }\gamma_3 \partial_{\mu}  \psi \; \\ \nonumber
& &  = 2k \chi_p \chi_a (\omega^1_{23}-\omega^2_{13}+\omega^3_{12}) + e^{0 \mu} (\chi_p \partial_{\mu} \chi_p + \chi_a \partial_{\mu} \chi_a)+ e^{3 \mu} (\chi_p \partial_{\mu} \chi_a - \chi_a \partial_{\mu} \chi_p) \eea

Finally, in order to stress the fact that vierbeins are viewed as fields, we will replace $e^{0\mu}$ through $e^{3 \mu}$ with $A^{\mu}$ through $D^{\mu}$ respectively, and introduce Lagrange multipliers to enforce orthonormality. We will also replace $\omega^1_{23}$ with $\omega^B_{CD}$ and do similarly with all the other indeces. These $\omega$-s are now functions of our vector fields that are defined based on formal substitution of these in place of Vierbeins without making an assumption of orthonormality, since the latter is only a consequence of Lagrange multipliers.  Thus, Lagrangian becomes 

\bea & & {\cal L}_{free}=2k \chi_p \chi_a (\omega^B_{CD}-\omega^C_{BD}+\omega^D_{BC}) + A^{ \mu} (\chi_p \partial_{\mu} \chi_p + \chi_a \partial_{\mu} \chi_a)+ D^{ \mu} (\chi_p \partial_{\mu} \chi_a - \chi_a \partial_{\mu} \chi_p) \; \nonumber \\
&  & +\ \lambda_1\,(A^\mu A_\mu-1) + \lambda_2\,(B^\mu B_\mu+1)
+ \lambda_3\,(C^\mu C_\mu+1) + \lambda_4\, (D^\mu D_\mu+1) \nonumber\\
& &\kern25pt+\ \lambda_5\,A^\mu B_\mu + \lambda_6\,A^\mu C_\mu
+ \lambda_7\, A^\mu D_\mu + \lambda_8\, B^\mu C_\mu
+ \lambda_9\, B^\mu D_\mu + \lambda_{10}\, C^\mu D_\mu \;,
\eea

where

\bea \omega^U_{VW} = U^{\mu} V^{\nu} (\partial_{\mu} W_{\nu} - \partial_{\nu} W_{\mu}) + V^{\rho} W^{\sigma} \partial_{\sigma} U_{\rho} \eea

\noindent{\bf 5. Interaction terms}

\noindent Now I would like to introduce interaction terms into Lagrangian. Since it is possible that we have interaction of more than one fermion, I would like to be able to define $\overline{\xi} \psi$ and $A_{\mu} \overline{\xi} \gamma^{\mu} \psi$.  In general, this means I would like to define $\overline{\xi} \Lambda \xi$ . Suppose vierbeins that are needed to put $\xi$ in the form $\chi_p u_1 + \chi_a v_1$ are $e_0^{\mu}=A^{\mu}$ through $e_3^{\mu}=D^{\mu}$ while vierbeins that are needed to put $\psi$ in the form $\eta_p u_1 + \eta_a  v_1$ are $f_0^{\mu}=E^{\mu}$ through $f_3^{\mu}=H^{\mu}$

Now, suppose the transformation from the $e$-basis to $f$-basis,  $e^{-1} f$ , lies in the connected component of identity matrix. In other words, they are either both forward-moving or both backward-moving. In either case, they are both forward-moving relative to each other. This means that we can write $e^{-1} f = \exp(\ln (e^{-1} f))$.  Thus, $\ln (e^{-1} f)$ can be viewed as generated by infinitesimal transformations. The infinitesimal spinor transformation that corresponds to $\ln (e^{-1} f)$ is $-\frac{\ii}{4} (\ln (e^{-1}f))_{\mu \nu}\, \sigma^{\mu \nu}$, where $\sigma^{\mu \nu} = \frac{\ii}{2} [ \gamma^{\mu}, \gamma^{\nu} ]$. Now, by exponentiating it back, we will get the finite spinor transformation corresponding to the transformation between these two coordinate systems:  $\exp\{-\frac{\ii}{4} (\ln (e^{-1} f))_{\mu \nu}\, \sigma^{\mu\nu}\}$. Thus, 

\bea \overline{\xi} \Lambda \psi =  (\chi_p \langle u_1 \vert + \chi_a \langle v_1 \vert)  \Lambda \exp\{-\frac{\ii}{4} (\ln (e^{-1} f))_{\mu \nu}\, \sigma^{\mu\nu}\} (\eta_p \vert u_1 \rangle + \eta_a \vert v_1 \rangle ) \eea

Now in the case where one reference frame is forward-moving and the other one is backward-moving, all we have to do is insert a time-reversal operator inside the log, namely $\exp\{-\frac{\ii}{4} (\ln(T L^{-1}M))_{\mu\nu} \sigma^{\mu\nu}\}$, where $T$ is time reversal. This gives us

\bea \overline{\xi} \Lambda \psi =  (\chi_p \langle u_1 \vert + \chi_a \langle v_1 \vert)  \Lambda T \exp\{-\frac{\ii}{4} (\ln (e^{-1} f))_{\mu \nu}\, \sigma^{\mu\nu}\} (\eta_p \vert u_1 \rangle + \eta_a \vert v_1 \rangle ) \eea

\noindent{\bf 6. Does Spin Mean Spinning?}

\noindent From the point of view of mathematical rigor, the equation we just got is as far as we can get. However, it would be fruitful to note that there is an intuitive correlation between that equation and the concept of ``spinning", which the word ``spin" represents. This discussion is not rigorous, and can be skipped by mathematically minded readers.

In order to visualize the ``spinning" that goes on, one can replace infinitesimal points in spacetime by small arrows. However, unlike the way it is normally done, these arrows will {\em not\/} be aligned with a spin axis. Instead, these arrows will be, themselves, spinning around some other axis. Thus, spin around the $z$ axis can be visualized as a vector pointing in the $x$ direction whose end is moving in the $y$ direction. Now, due to the fact that spin is subject to Lorentz transformations, we also need to know the reference frame, and it is given by $A^{\mu}$. And finally $D^{\mu}$ is a cross product of $B^{\mu}$ with $C^{\mu}$ in a reference frame in which the particle is at rest. Thus $D^{\mu}$ is what is usually thought of as the direction of spin.

Now let's look at each term in the Lagrangian to see what it represents. First, consider the $C^{\mu} A^{\nu} \partial_{\nu} B_{\mu}$ term. In the reference frame of the point around which the arrow spins, $A^{\nu}=\delta^{\nu}{}_0$ is just a vector pointing along the $t$ axis. Thus, in this reference frame, the term becomes $C^{\mu} \partial_0 B_{\mu}$. Now, if we visualize $B^{\mu}$ as pointing along the $x$ axis and $C^{\mu}$ along the $y$ axis, this expression reads off as ``how fast does the end of the $x$ axis move in the $y$ direction"? Note that the end of the $x$ axis can also move in the $z$ direction, but this speed would simply have no contribution to the Lagrangian. In other words, the way to think of it is this: $A^{\mu}$, $B^{\mu}$ and $C^{\mu}$ are vector fields with a weird coupling between them. They are coupled to each other in such a way that the end of vector $B^{\mu}$ is ``forced" to move in the direction of the vector $C^{\mu}$ by the Lagrangian, which would then be interpreted by the observer as a rotation around $D^{\mu}$.

Now let's look at the $A^{\mu} B^{\nu} \partial_{\nu} C_{\mu}$ and $A^{\mu} B^{\nu} \partial_{\nu} C_{\mu}$ terms. One difference between the first term and these two terms is that in the ``directional derivative" part of the equation (which in the first term is $A^{\nu} \partial_{\nu}$) $A^{\nu}$ is being replaced by $B^{\nu}$ and $C^{\nu}$, respectively.  This means that the differentiation is no longer in the time direction, but rather in a spatial direction. The interpretation of this might be that, as opposed to speaking of the rotation of one arrow (in which case it travels along the $y$ direction) we are comparing the motions of different arrows (and the spatial direction differentiates the arrows we are looking at). The other difference is that while the first term refers to angular motion, both of the other two terms refer to linear motion: $A^{\mu} B^{\nu} \partial_\nu C_\mu$ refers to ``boosting" of the spatial axis $C_{\mu}$ in the time direction $A^\mu$ (which would be proportional to the {\em negative\/} velocity in the $C_\mu$ direction), while $B^\mu C^{\nu} \partial_\nu A_\mu$ refers to ``boosting" of the time axis $A_\mu$ in the spatial direction $B^\mu$ (which can be interpreted as a {\em positive\/} velocity in the $B_\mu$ direction). Thus, by identifying $B^{\mu}$ with the $x$ axis and $C^\mu$ with the $y$ axis, the Lagrangian tells us that we ``want" points that are located away along the $x$ direction to move in the negative $y$ direction, and we want points located away along the $y$ direction to move in the positive $x$ direction. This is equivalent to saying that we want to have a ``planetary system" and we want the orbits of the planets to have spin in the $-z$ direction.

This can be summarized as follows: we can envision space to be constructed of mini-atoms.  The first term in our Lagrangian tells us about the spin of each electron in an atom, while the last two terms tell us about the orbital rotation of electrons. However, these atoms are ``glued together" so that the orbital rotation of one atom gets ``passed" onto neighboring ones, which is why this actually looks like a derivative globally. At the same time, while there is ``cohesion" between atoms in the second two terms, the first term has no such thing: the arrow is infinitesimal and doesn't extend to a neighboring point. Similarly, when we talk about linear motion in the last two terms, this linear motion is really a similar arrow pointing in the $t$ direction, which is also infinitesimal. Thus, while there {\em is\/} spatial cohesion in the second two terms, there is no time cohesion.

\noindent{\bf 7 . Conclusion}

\noindent As I have shown in this paper, it is possible to completely define fermions by using scalar and vector fields. I was able to both define inner products of fermions as well as their Lagrangians by simply referring to the relevant vector and scalar fields, with a redefined set of transformation laws.

This approach has several benefits: (1) Philosophical ones, including the fact that it is now much easier to visualize fermions the way we visualize all other fields, and that there is no more ``weirdness" about the $z$ axis being ``special" when we talk about ``spin-up", since now the $z$ axis is being replaced by a vector field $D^{\mu}$, and these other fields are special because they are coupled to each other in a certain way. (2) Since we have equated fermions with reference frames, it can be argued that the manifold-like topology is not ÒinnateÓ but rather is the result of a fermionic field ``organizing" its surroundings at each point into local coordinates. In particular, this will be useful in providing a mechanism of causal sets becoming manifoldlike without aforegiven coordinate system. A more detailed discussion of this point will be included in Ref \cite{spinors}.


\end{document}